\documentclass[a4paper]{article}
\usepackage{amsmath}
\usepackage{amssymb}
\usepackage{graphicx}
\usepackage{fullpage}
\usepackage{url}

\newcommand{\R}{\mathbb{R}}
\newcommand{\Dmn}{D_{n}^{m}}
\newcommand{\Dm}{D^{m}}
\newcommand{\SEQ}{\mathcal{S}}
\newcommand{\s}{^{\ast}}

\title{The Hadwiger Number of Jordan Regions is Unbounded%
  \thanks{This research was supported by the Korea Research Foundation.}}

\author{Otfried Cheong%
  \thanks{Division of Computer Science, Korea Advanced Institute of
    Science and Technology, Daejeon, South Korea.
    Email:~\{otfried, mira\}@tclab.kaist.ac.kr.}
  \and Mira Lee%
  \footnotemark[2]}

\date{}

\newtheorem{theorem}{Theorem}
\newtheorem{lemma}{Lemma}

\makeatletter
\newbox\ProofSym
\setbox\ProofSym=\hbox{%
\unitlength=0.18ex%
\begin{picture}(10,10)
\put(0,0){\framebox(9,9){}}
\put(0,3){\framebox(6,6){}}
\end{picture}}
\newenvironment{proof}[1][Proof.]{\O@proof{#1}}{\O@endproof}
\def\O@proof#1{\trivlist\@topsep\z@\@topsepadd\smallskipamount
  \@ifstar{\item[]}{\item[\hskip\labelsep\it #1 ]}}
\def\O@endproof{\hfill\copy\ProofSym\linebreak\endtrivlist}
\makeatother

\let\geq\geqslant
\let\leq\leqslant

\makeatletter
\partopsep\z@ \textfloatsep 10pt plus 1pt minus 4pt
\def\section{\@startsection {section}{1}{\z@}{-3.5ex plus -1ex minus
-.2ex}{2.3ex plus .2ex}{\large\bf}}
\def\subsection{\@startsection{subsection}{2}{\z@}{-3.25ex plus -1ex
minus -.2ex}{1.5ex plus .2ex}{\normalsize\bf}}
\def\@fnsymbol#1{\ensuremath{\ifcase#1\or *\or **\or 1\or 2\or 3\or 4\or 5\or
    6\or 7\or 8\or 9\else\@ctrerr\fi}}
\makeatother

\begin{document}
\maketitle

\begin{abstract}
  We show that for every $n > 0$ there is a planar topological disk
  $A_{0}$ and $n$~translates $A_{1}, A_{2}, \dots, A_{n}$ of $A_{0}$
  such that the interiors of $A_{0}, \dots A_{n}$ are pairwise
  disjoint, but with each $A_{i}$ touching~$A_{0}$ for~$1 \leq i \leq
  n$.
\end{abstract}

\section{Introduction}

For any compact body $C \subset \R^{d}$, we define $H(C)$, the
Hadwiger number of $C$, as the maximum number of mutually
non-overlapping translates of $C$ that can be brought into contact
with~$C$ (see the survey by Zhong~\cite{z-kncb-98}).
Hadwiger~\cite{h-utte-57} showed that for convex sets $C$ we have
$H(C) \leq 3^{d} - 1$ (using Minkowski's difference body method, see
also Gr\"unbaum~\cite{g-ch-61}). The bound is tight for
parallelepipeds~\cite{g-aakk-61, g-ch-61}. In the planar case it is
known that $H(C) = 6$ for every convex $C$ other than a parallelogram.

The arguments used in these results rely strongly on convexity.
Considering the more general family of \emph{Jordan
regions}\footnote{A set $C \subset \R^{2}$ is a \emph{Jordan region}
or \emph{topological disk} if it is bounded by a closed Jordan curve,
or equivalently if it is homeomorphic to the unit disk.} in the plane,
Halberg et al.~\cite{hls-ccses-59} could show that $H(C) \geq 6$ holds
for any Jordan region $C \subset \R^{2}$.  More precisely, they showed
that there exist six non-overlapping translates of~$C$ all
touching~$C$ and whose union \emph{encloses} $C$, where a set $A$
encloses a set $B$ if every unbounded connected set which intersects
$B$ also intersects~$A$.  It seems therefore natural to conjecture
that the Hadwiger numbers of Jordan regions in the plane are bounded
by an absolute constant.  Some more evidence for this conjecture is a
result of Bezdek et al.~\cite{bkk-mctpd-95} who showed that the
maximum number of pairwise touching translates of a Jordan region $C$
in the plane is four.  Since in this respect Jordan regions behave in
the same way as convex sets, they ask the following
question:
\begin{quote}
  It seems reasonable to conjecture that $H(C) \leq 8$ for every
  planar Jordan region~$C$.  If this conjecture is false, is there an
  upper bound for $H(C)$ independent from the disk~$C$?\\
  (\emph{A.~Bezdek, K.~and W.~Kuperberg~\cite{bkk-mctpd-95}; Problem~6.1})
\end{quote}
As a first step in settling this conjecture, A.~Bezdek could later
show that $H(C) \leq 75$ if $C$~is a \emph{star-shaped} Jordan region.
The problem was picked up again by Brass et al.~\cite{bmp-rpdg-05}
(Problem~5 and Conjecture~6 in Section~2.4).  We show here that the
conjecture is not true in a strong sense: The Hadwiger number of
Jordan regions in the plane is not bounded by \emph{any} constant.
For each $n > 0$, we construct a Jordan region that admits $n$
mutually non-overlapping translates touching it.

The case of star-shaped Jordan regions remains open in the weaker
sense of establishing the right constant: Brass et al.~conjecture that
this constant is eight, but the best known upper bound is~$75$.

\section{The proof}

We consider the integer sequence $\SEQ = s_{1},s_{2},\dots$ where $s_{i}$
is the number of bits that must be counted from
right to left to reach the first $1$ in the binary representation of~$i$.
\begin{align*}
  \SEQ & =  1, 2, 1, 3, 1, 2, 1, 4, 1, 2, 1, 3, 1, 2, 1, 5, 1, 2, 1,
  3, 1, 2, 1, 4, 1, 2, 1, 3, 1, 2, 1, 6 \dots
\end{align*}
This sequence, which is also known as the ruler function, appears as
sequence \emph{A001511} in the on-line encyclopedia of integer
sequences.\footnote{\url{http://www.research.att.com/~njas/sequences/A001511}}
We need the following property of this sequence.
\begin{lemma}
  \label{lem:prefix-sum}
  The prefix of length~$k$ of $\SEQ$ has the smallest sum among all
  subsequences of length~$k$ of $\SEQ$, for any $k > 0$.  Formally,
  for any $k,r > 0$
  \[
  \sum_{i = 1}^{k} s_{i} \leq \sum_{i = r}^{r+k-1} s_{i}.
  \]
\end{lemma}
\begin{proof}
  We proceed by induction.  If $k = 1$, the claim is true since $s_{1}
  = 1 \leq s_{r}$.  Assume now that $k > 1$ and that the claim is true
  for all shorter prefixes.  If $k$ is odd, then $s_{k} = 1$, and by
  induction we have
  \[
  \sum_{i = 1}^{k} s_{i} = 1 + \sum_{i = 1}^{k-1} s_{i}
  \leq 1 + \sum_{i = r}^{r + k - 2} s_{i} \leq
  \sum_{i = r}^{r + k - 1} s_{i}.
  \]
  It remains to consider even $k$.  We observe that every odd term of
  $\SEQ$ is equal to~$1$, and that $\SEQ$ has a nice recursive structure:
  removing all odd terms and subtracting one from all even terms
  results in the same sequence~$\SEQ$ again.  We therefore have
  \[
  \sum_{i=1}^{k} s_{i} =
  k/2 + \sum_{i=1}^{k/2} (s_{i} + 1)
  = k + \sum_{i=1}^{k/2} s_{i}
  \leq k + \sum_{i=r'}^{r' + k/2 - 1} s_{i}
  = k/2 + \sum_{i=r'}^{r' + k/2 - 1} (s_{i} + 1)
  = \sum_{i=r}^{r+k-1} s_{i},
  \]
  where $r' = \lceil r/2\rceil$.
\end{proof}

We can now describe our planar topological disk, or more precisely, a
two-parameter family of disks.  For integers $m \geq 2$ and $n \geq
1$, the disk~$\Dmn$ is the union of $2^{n}$ horizontal \emph{bars}
$B_{1},\dots, B_{2^{n}}$ and $2^{n}-1$ vertical \emph{connectors}
$V_{1},\dots, V_{2^{n} -1}$.  All bars are axis-parallel rectangles of
width~$m$ and height~$1$, all connectors are axis-parallel rectangles
of width~$1$. The height of connector $V_{i}$ is~$s_{i}$ (the $i$th
term of our sequence~$\SEQ$).  Informally, connector $V_{i}$ is placed
above the rightmost unit square of bar~$B_{i}$, while bar $B_{i+1}$ is
placed to the right of the topmost unit square of connector~$V_{i-1}$.

Formally, bar $B_{i}$ is the rectangle spanning the $x$-interval
$[(i-1)m, im]$ and the $y$-interval $[y_{i},y_{i}+1]$, where $y_{i} =
\sum_{j = 1}^{i-1} s_{j}$.  Connector $V_{i}$ spans the $x$-interval
$[im-1,im]$ and the $y$-interval $[y_{i}+1,y_{i+1}+1]$.

\begin{figure}[ht]
  \centerline{\includegraphics[width=\textwidth]{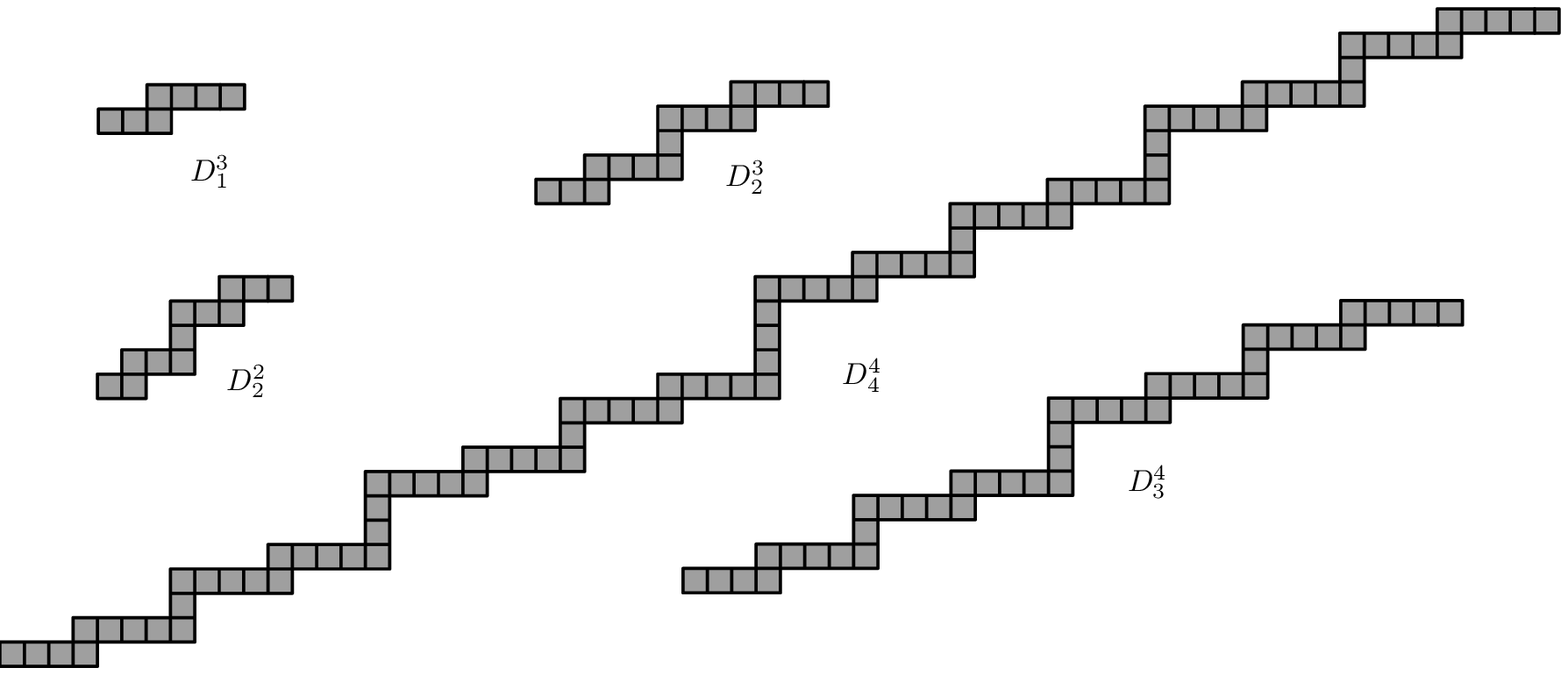}}
  \caption{Some $\Dmn$.}
  \label{fig:examples}
\end{figure}

Figure~\ref{fig:examples} shows $\Dmn$ for some values of $m$ and $n$.

We can give an alternative, recursive description of $\Dm_{n}$, by
observing that bars $B_{1},\dots, B_{2^{n-1}}$ and bars
$B_{2^{n-1}+1},\dots, B_{2^{n}}$ of $\Dmn$ form two translates of
$\Dm_{n-1}$, connected by the single connector~$V_{2^{n-1}}$. We can
consider $\Dmn$ to consist of two translates of $\Dm_{n-1}$, or four
translates of $\Dm_{n-2}$, or $2^{n-1}$ translates of $\Dm_{1}$, or,
in general, $2^{k}$ translates of $\Dm_{n-k}$.
\begin{lemma}
  \label{lem:disjoint-steps}
  Let $A$ and $A'$ be translates of $\Dmn$, for $m, n \geq 2$, such that
  the first bar $B_{1}'$ of $A'$ is obtained from some bar $B_{r}$ of
  $A$ by a translation of $y\s \geq 1$ downwards and $1 \leq x\s \leq m-1$
  to the right, where $1\leq r \leq 2^{n}$.  Then $A$ and $A'$ have
  disjoint interior.
\end{lemma}
\begin{proof}
  Consider the vertical strip spanned by bar $B_{r-1+i}$ of~$A$, for
  $1 \leq i \leq 2^{n} - r + 1$.  Since $1 \leq x\s \leq m-1$, this
  strip can intersect only bars $B'_{i-1}$ and $B'_{i}$ and connector
  $V'_{i-1}$ of~$A'$.  The highest $y$-coordinate in $B'_{i-1} \cup
  V'_{i-1} \cup B'_{i}$ is $y_{i}+1$ with respect to the origin
  of~$A'$.  By assumption, the origin of $A'$ is at $y$-coordinate
  $y_{r} - y\s\leq y_{r} -1$, and so $B'_{i-1} \cup V'_{i-1} \cup
  B'_{i}$ lies below the line $y = y_{r} + y_{i}$.  On the other hand,
  the bottom edge of $B_{r-1+i}$ of~$A$ has
  $y$-coordinate~$y_{r-1+i}$.  We now have
  \[
  y_{r-1+i} - (y_{r} + y_{i}) 
  = ( y_{r-1+i} - y_{r} ) - y_{i}
  = \sum_{j=r}^{r+i-2}s_{j} - \sum_{j=1}^{i-1}s_{j} 
  \geq 0
  \]
  by Lemma~\ref{lem:prefix-sum}.  This implies that the interior of
  $B'_{i-1} \cup V'_{i-1} \cup B'_{i}$ lies stricly below $B_{r-1+i}$,
  and the lemma follows.
\end{proof}

We can now describe our construction of touching translates.  We fix
an integer $n > 1$, and pick any integer $m \geq n$.  Let $A_{1}$
be~$\Dmn$.  For $2 \leq i \leq n$ we obtain $A_{i}$ from $A_{i-1}$
as follows: The \emph{first} (leftmost) copy of $\Dm_{n+1-i}$ in
$A_{i}$ is a translate of the \emph{second} copy of $\Dm_{n+1-i}$ in
$A_{i-1}$, translated down by one and right by one.

We observe now that for any pair $1 \leq i < j \leq n$, the leftmost
copy of $\Dm_{n+1-j}$ in $A_{j}$ is a translate of some copy of
$\Dm_{n+1-j}$ in $A_{i}$, translated down by $j-i$ and right by~$j-i$.
Since $1\leq j-i < n \leq m$, Lemma~\ref{lem:disjoint-steps} implies
that the interiors of $A_{i}$ and $A_{j}$ are disjoint.

\begin{figure}[ht]
  \centerline{\includegraphics{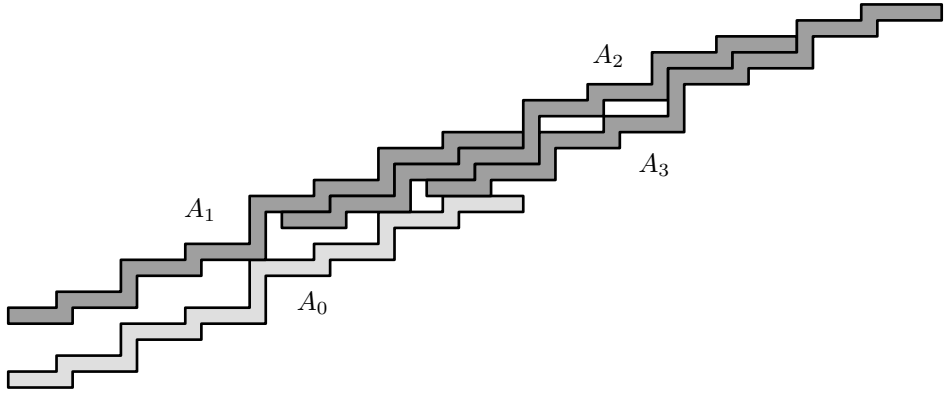}}
  \caption{The construction for $m = 4$, $n = 3$.}
  \label{fig:cons-4-3}
\end{figure}

Let now $A_{0}$ be a translate of $A_{1}$, translated downwards
by~$n+1$.   See Figure~\ref{fig:cons-4-3} for the entire construction
for $m = 4$, $n = 3$.

It remains to show that $A_{i}$ touches $A_{0}$, but that their
interiors are disjoint, for $1 \leq i \leq n$.

We pick some $i$.  Let $D$ be the \emph{last} (rightmost) copy of
$\Dm_{n+1-i}$ in $A_{0}$.  Then the first copy $D'$ of $\Dm_{n+1-i}$
in $A_{i}$ can be obtained from $D$ by translating upwards by $n+1$,
then downwards by $i-1$ and right by~$i-1$. In other words, $D'$ is
obtained from $D$ by translating upwards by $n+2-i$ and right by
$i-1$.  Now, the middle vertical segment of $D$ is a rectangle of
height~$n+2-i$, and so this translation brings $D$ and $D'$ into
contact.  On the other hand, all other vertical segments in $D$ have
length less than $n+2-i$, and so the interiors of $D$ and $D'$ are
disjoint.  Finally, since $D$ is the rightmost part of $A_{0}$ and
$D'$ is the leftmost part of $A_{i}$, no other intersections between
$A_{0}$ and $A_{i}$ are possible, and so their interiors are disjoint.

\begin{figure}[ht]
  \centerline{\includegraphics[width=\textwidth]{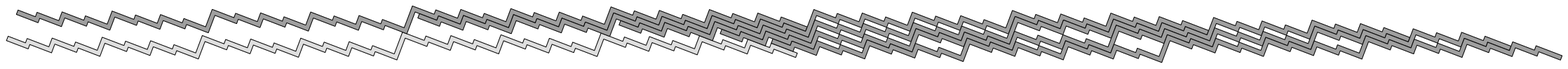}}
  \caption{The construction for $m = n = 5$.}
  \label{fig:cons-5-5}
\end{figure}

We summarize the result in the following
theorem. Figure~\ref{fig:cons-5-5} shows a larger example.
\begin{theorem}
  For any integer $n \geq 2$ and any integer $m \geq n$ there are $n+1$
  translates $A_{0}, A_{1},\dots, A_{n}$ of $\Dmn$ whose interiors are
  pairwise disjoint, but such that $A_{0}$ touches every $A_{i}$, $1
  \leq i \leq n$.
\end{theorem}

\bibliographystyle{plain}
\bibliography{biblio}
\end{document}